# A Thematic Framework for Analyzing Large-scale Self-reported Social Media Data on Opioid Use Disorder Treatment Using Buprenorphine Product

Madhusudan Basak[1,8], Omar Sharif[1], Sarah E. Lord[2,3,6], Jacob T. Borodovsky[2,3], Lisa A. Marsch[2,4], Sandra A. Springer[9,10], Edward Nunes[11], Charlie D. Brackett[5,7], Luke J. ArchiBald[4,6], Sarah M. Preum[1,2,3]

[1]Department of Computer Science, Dartmouth College
[2]Center for Technology and Behavioral Health, Geisel School of Medicine, Dartmouth College
[3]Department of Biomedical Data Science, Geisel School of Medicine, Dartmouth College
[4]Department of Psychiatry, Geisel School of Medicine, Dartmouth College
[5]Department of Medicine, Geisel School of Medicine, Dartmouth College
[6]Department of Psychiatry, Dartmouth Health
[7]Department of Internal Medicine, Dartmouth Hitchcock Medical Center
[8]Department of CSE, Bangladesh University of Engineering and Technology
[9]Department of Internal Medicine, Yale School of Medicine
[10]Center for Interdisciplinary Research on AIDS, Yale University School of Public Health
[11]Department of Psychiatry, Division on Substance Use, Columbia University Irving Medical Center and New York State Psychiatric Institute
{madhusudan.basak.gr, sarah.masud.preum}@dartmouth.edu

## Abstract

**Background:** One of the key FDA-approved medications for Opioid Use Disorder (OUD) is buprenorphine. Despite its popularity, individuals often report various information needs regarding buprenorphine treatment on social media platforms like Reddit. However, the key challenge is to characterize these needs. In this study, we propose a theme-based framework to curate and analyze large-scale data from social media to characterize self-reported treatment information needs (TINs).

**Methods**: We collected 15,253 posts from r/Suboxone, one of the largest Reddit sub-community for buprenorphine products. Following the standard protocol, we first identified and defined five main themes from the data and then coded 6,000 posts based on these themes, where one post can be labeled with applicable one to three themes. Finally, we determined the most frequently appearing sub-themes (topics) for each theme by analyzing samples from each group.

**Results**: Among the 6,000 posts, 40.3% contained a single theme, 36% two themes, and 13.9% three themes. The most frequent topics for each theme or theme combination came with several key findings - prevalent reporting of psychological and physical effects during recovery, complexities in accessing buprenorphine, and significant information gaps regarding medication administration, tapering, and usage of substances during different stages of recovery. Moreover, self-treatment strategies and peer-driven advice reveal valuable insights and potential misconceptions.

**Conclusions:** The findings obtained using our proposed framework can inform better patient education and patient-provider communication, design systematic interventions to address treatment-related misconceptions and rumors, and streamline the generation of hypotheses for future research.

# 1. Introduction

Opioid Use Disorder (OUD) remains a significant public health concern in the United States (US). More than 81,000 deaths from opioid overdoses were reported in 2022, contributing to nearly 7,25,000 deaths from 1999 to 2021 (Wide-ranging online data for epidemiologic research: (WONDER), 2024). buprenorphine, methadone, and naltrexone are FDA-approved medications for opioid use disorder (MOUD) and the gold-standard treatments for OUD (Livingston et al., 2022; Pizzicato et al., 2020; Yarborough et al., 2016). However, individuals considering or undergoing MOUD treatment often report a range of information needs related to different aspects of treatment (Mackey et al., 2020), including accessing MOUD, medication schedule (timing, dosage), concurrent substance use, unexpected symptoms and side effects, and tapering off MOUD. When unaddressed, these issues can result in non-compliance with treatment, causing delays, discontinuation, or resorting to unverified treatments (Marks et al., 2023; Yarborough et al., 2016).

**Identifying and characterizing the treatment information needs (TINs)** of individuals with OUD is a critical first step to designing effective interventions for MOUD treatment induction, adherence, and retention. A proper analysis in this regard involves curating a suitable dataset with TINs, classifying the data into granular categories (herein we referred to those as *themes*), and finding out theme-based characteristics, i.e., how these themes co-occur, the sub-themes or patterns that appear more frequently, and how individuals address their TINs.

**Social media, particularly anonymous platforms like Reddit,** can help us capture the diversity of TINs of thousands of individuals with lived experiences and the real-world complexity of recovery (Chen and Wang, 2021; Edo-Osagie et al., 2020; Jha and Singh, 2020; Pandrekar et al., 2018; Paul et al., 2016; Skaik and Inkpen, 2021). In the US, approximately 70% of the population uses social media to share information with their peers (Kanchan and Gaidhane, 2023). What sets social media apart from traditional data sources is the spontaneous, self-reported lived experiences shared by individuals, a type of data that is not easily obtainable through other data sources like electronic health records (EHR) or surveys. This aspect is particularly vital in the context of OUD, a highly stigmatized topic where patients often hesitate to reach out to traditional healthcare providers to address treatment information needs due to a lack of access, trust, or health equity (Nobles et al., 2021).

**The current research on identifying MOUD-related TINs on social media** has two major limitations. **First**, there exists no publicly available, labeled, large dataset (i.e., dataset fully curated by knowledgeable coders and verified by experts) that covers self-narrated discourse on MOUD TINs from over three thousand affected individuals. The majority of the existing datasets are small in size, which effectively restricts the generalizability of the findings. **Second**, existing works mainly consider only one theme/type of TINs for a MOUD treatment option, e.g., logistical barriers to accessing the treatment or the physical or psychological effects of the treatment. Collecting and analyzing data considering only a single theme can lead to a loss of associated valuable correlations and nuanced insights. To comprehensively understand interconnected issues and the overall scenario, it is crucial to analyze posts encompassing multiple themes.

**We address the research gap by proposing a novel theme-driven framework** that provides a labeled dataset, and a comprehensive analysis is performed on this dataset, covering multiple themes of TINs and capturing data from thousands of affected individuals. This framework is built upon existing research and can complement and augment focus groups or other relevant methods. As a use case of MOUD, we considered buprenorphine products, as they are one of the most widely used and available MOUD treatment options (Shulman et al., 2019). We implemented our framework on r/Suboxone, one of the most popular Reddit communities for buprenorphine products, with about 39,000 members to date. We collected 15,253 posts from this subreddit between January 2, 2018, and August 6, 2022, and randomly selected 6,000 posts generated by 3,372 unique individuals. Then, we labeled/coded the 6,000 posts according to different themes of TINs. To our knowledge, this is the largest dataset on patient-reported TINs on buprenorphine products for OUD treatment. We conducted a thorough thematic analysis of the labeled data to surface several key insights on the challenges the individuals face - their information needs and knowledge gaps regarding treatment initiation or continuation with buprenorphine, their experiences at different stages of recovery, self-treatment strategies, and treatment misperceptions.

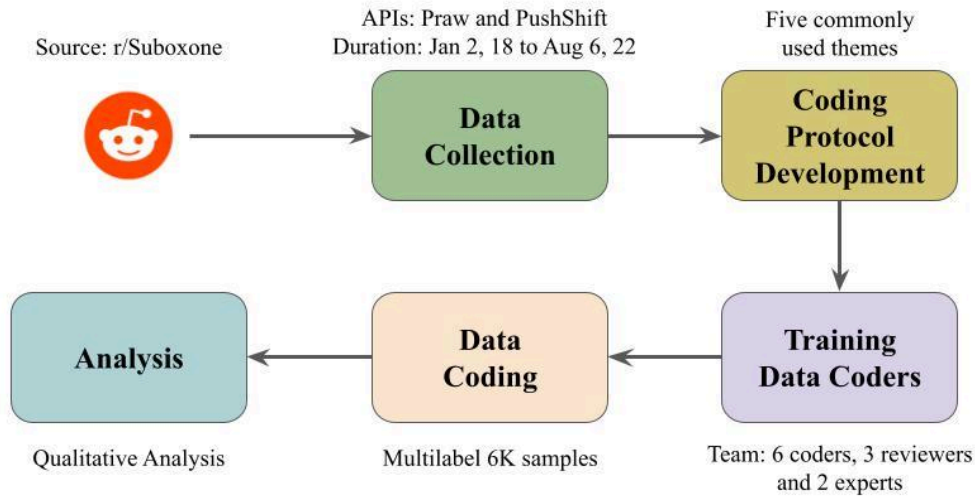

Figure 1: Methodology Flow Diagram

## 2. Methods

We used a comprehensive methodological framework, encompassing the entire process from selecting data sources to conducting experimental analyses. Figure 1 highlights the key steps of this process. This study received approval from the Institutional Review Board (IRB) at the authors' institution.

**2.1 Data Collection**
Using PRAW and PushShift APIs (PRAW: The Python Reddit API Wrapper, 2023; Pushshift Reddit API Documentation, 2023), we collected posts, comments, likes, upvotes, and unique post identifiers from r/Suboxone, the largest Reddit community dedicated to Suboxone (individuals also often discuss other buprenorphine products) with over 39,000 members to date. We followed Reddit policy and institutional IRB protocol to collect and post-process data. Subsequently, we had 15,253 posts available for our study. Due to resource constraints, we randomly chose 6,000 posts for manual coding, where each post contained less than 300 words. The length restriction was implemented to ensure a more focused and precise assessment by the coders.

**2.2 Develop Data Coding Protocol**
After collecting the data, we developed a data coding protocol for identifying themes in TINs, which followed the following steps.

**Step 1: Develop the initial coding protocol.** Following the standard qualitative method, the authors used an iterative coding process to identify and delineate the themes that emerge from data in a naturalist way. At first, authors SP, MB, and OS employed an inductive approach on a subset of 250 samples and identified the recurring themes from the data. They also documented the initial definition of these themes. Then, they shared the theme definitions with a relevant subset of examples with experts SEL and JB. SEL is an expert in clinical psychology in practice and research for populations receiving MOUD, while JB is an expert in social media platforms for buprenorphine treatment. The experts (SEL and JB) deployed an abductive analysis method. They complemented the findings from the inductive analysis with deductive analysis using existing theories or hypotheses related to information-seeking behavior and information needs of individuals considering or undergoing MOUD treatment. Based on the expert feedback, the initial themes were consolidated into five main themes. The authors followed the thematic saturation method to build consensus around the primary themes. Then, these five authors (SP, MB, OS, SEL, JB) developed a systematic coding protocol to define and delineate the five main themes of TINs.

**Step 2: Revise coding protocol through iterative data coding.** Then, the five authors iteratively coded an additional 1,250 posts to define the scope and boundaries of each theme and revise the coding protocol to delineate each theme concretely. In this phase, the authors found some samples that neither fall under the five primary themes nor show a pattern to call for a new theme. Hence, these outlier samples were categorized as having a generic "Others" theme. In addition, the whole process also generated supplementary information for coding (e.g., a dictionary of different brand names for buprenorphine products, street names, and variations of different substances, as they are often mentioned in Reddit posts).

**Step 3: Triangulation.** To ensure the consistency and reliability of the coding protocol, we applied researcher triangulation. Specifically, three additional subject matter experts (LM, SS, and EN) conducted an unbiased review of the coding protocol and annotated data samples from each theme. All of these experts are addiction researchers, and two of them also practice addiction psychiatry. Specifically, they were asked to review the correctness of theme definition and labeling, any missing critical themes, and any issues that need clarification. The coding protocol was updated based on their feedback and suggestions.

Table 1 describes the resulting themes related to buprenorphine product-specific TINs. Some themes reflect the early stage of recovery (e.g., AccBup, TaekBup), while others reflect the recovery continuum (e.g., CoSU, Psyphy).

*Table 1: Definitions of the main themes delineated by the experts are presented here. The shorthands and themes have been mentioned. These shorthands have been used throughout the rest of the article.*

| Theme Name | Description |
|---|---|
| **Accessing buprenorphine (AccBup)** | This theme addresses concerns about accessing buprenorphine, e.g., challenges with insurance, pharmacies, and healthcare providers. Identifying these barriers can help understand factors that affect treatment initiation, adherence, and retention. |
| **Taking buprenorphine (TakeBup)** | This category highlights concerns about the treatment regimen for buprenorphine, e.g., questions on dosage, timing, and frequency. It emphasizes the potential for misconceptions that may hinder treatment adherence. |
| **Experiencing Psychophysical Effects during Recovery (Psyphy)** | This theme includes concerns about the physical and psychological effects experienced or anticipated during recovery. It covers both rare and common effects of buprenorphine, examining how these may influence treatment adherence. |

| Theme Name | Description |
|---|---|
| **Co-occurring Substance usage (CoSU)** | This theme explores concerns related to using other substances during recovery, whether for recreational use or self-medication. It offers insights into individuals' experiences with substance use alongside buprenorphine treatment. |
| **Tapering buprenorphine (TapeBup)** | This theme focuses on concerns about reducing or discontinuing buprenorphine use. It provides insights into self-tapering practices, including reasons, timing, and the effectiveness of tapering strategies. |

**2.3 Large-scale Data Coding**

We performed a three-month-long rigorous large-scale coding of an additional 4,500 samples by a team of eleven members. Three authors (MB, OS, SP) formed the reviewer team, and two seasoned addiction researchers and faculty members (SEL and JB) were in the expert team. The coding team comprised three graduate students and three undergraduates recruited at the authors' institution. The coding team underwent rigorous training sessions led by the reviewer team to gain subject matter knowledge. The coders performed four rounds of coding spanning over six weeks. To maintain the quality of our coding, each sample was independently assigned to two different coders. The reviewer team resolved the disagreements and finalized the coding. Finally, the experts independently reviewed 100 samples from the dataset to ensure the quality and consistency of the final dataset.

**2.5 Determining Topics within Themes**

To determine the most frequent topic(s)/sub-themes for each theme, we randomly selected at most 50 posts from each theme category. This choice of 50 posts was made as it represents roughly one-fifth of the average number of posts (240 posts) within each theme. For each theme, two coders (graduate students) assigned topic(s) to each selected post independently. Consequently, they discussed themselves, came across a unique list of topics for that theme, and revised their initial topic assignment to the posts. Finally, they identified and reported the most frequent topic(s) for each theme.

## 3. Results

**3.1 Theme-wise Data Distribution**

In our dataset, we labeled each post with themes evoked from the data. Each theme either appeared as a standalone theme or co-occurred with other themes in a post. We observed some posts cover up to three different themes. So, we break down the frequency distribution into three groups – posts labeled with one theme, followed by posts with two themes, and posts with three themes. Table 3 presents the distribution of posts among different groups. Additionally, 588 posts (9.8% of the entire dataset) were categorized as *Others (Oth)*, indicating that they did not have any themes as a label.

Among the total 6,000 posts, 2,417 (40.3%) were labeled with only <u>one theme</u> (Rows 1-5). The frequency distribution of individually labeled themes was uneven, with *Psyphy* being the most common (700 posts), followed by *AccBup* (672 posts). *TapeBup* had the least frequent occurrence, observed in 237 posts.

We identified 2160 posts (36%) with <u>two themes</u> (Rows 6-15). *Psyphy-TapeBup* (Posts labeled with both *Psyphy* and *TapeBup)* was the most common, appearing in 738 posts. In contrast, only 23 posts were labeled with *AccBup-TapeBup*. Conversely, 835 posts (13.9%) were tagged with three themes (Rows 16-25). <u>Three</u> combinations dominated: a) *CoSU-TakeBup-Psyphy* (312 posts), b) *CoSU-Psyphy-TapeBup* (233 posts), and c) *TakeBup-Psyphy-TapBup* (142 posts). The remaining seven combinations contributed to a total of 148 posts.

### 3.2 Findings Surfaced from Thematic Analysis

**Prevalence of reporting psychophysical effects during recovery**: *Psyphy* is a highly prevalent theme, occurring individually or with other themes. We analyzed the data to separately identify (i) the common *physical* and *psychological* effects within self-reported contents of the posts and (ii) the potential correlation with other themes. We randomly sampled 100 posts from the pool of posts tagged with *Psyphy* as one of the themes, i.e., *Psyphy* as a standalone theme and any combination of themes that includes *Psyphy*. Based on qualitative analysis, we found that the psychological effects (e.g., *anxiety, suicidal thoughts, anger*) are often correlated with themes of *Co-occurring Substance usage (CoSU)* and *Taking buprenorphine (TakeBup)*. The scenario is different for the physical effects. Although physical effects are common in all combinations, each has a separate list of physical effects. For example, *puking, sweating, restless leg*s, etc., are common when theme *Psyphy* co-occurred with theme *TakeBup*, whereas *insomnia, low energy, sneezing, etc.,* are available when theme *Psyphy* co-occurred with theme *Tapering buprenorphine (TapeBup)*. Unsurprisingly, the only exception is *withdrawal*, which is common in almost every combination, i.e., whether theme *Psyphy* occurs as a standalone theme or co-occurs with any other theme. These findings have implications for clinical and public health research as well as targeted interventions for patient communication and education.

**Complexities stemming from barriers to accessing buprenorphine (AccBup):** Our thematic analysis reveals different contexts of access barriers to buprenorphine as well as a wide range of complexities stemming from lack of access to buprenorphine (*AccBup*) that can complement existing research on improving access to treatment. For example, row 2 of Table 4 shows examples of individuals asking questions to resolve different access barriers, including accessing telehealth options for treatment, issues with prescription refills at the pharmacy, and insurance coverage. While analyzing samples labeled with other themes in addition to *AccBup*, we find several complexities stemming from the lack of access to treatment. Such as considering other buprenorphine formulations (e.g., Suboxone to Zubsolv in row 14), tapering and self-dosing buprenorphine (row 15) and self-treatment strategies to manage the associated psychophysical effects (rows 19, 22, 23), and considering other substances (e.g., *Oxycodone, Kratom, Hydrocodone*) to reduce the side effects caused by stopping or lowering the dose of buprenorphine product (rows 21, 24, 25). Further analysis of such discourse can improve the understanding of stigma and knowledge gaps associated with recovery treatment in a community-informed way.

**Information gaps regarding taking of buprenorphine products (TakeBup):** While analyzing single and co-occurring instances of the theme *TakeBup*, we identified several cases of information gaps related to taking buprenorphine that are prevalent in online communities discussing recovery treatment using buprenorphine products. These include techniques to administer (e.g., dissolving under the tongue, spitting, swallowing) the buprenorphine products, absorption rate, ease of use, dosing, brand comparison (row 3), how to dose during tapering (row 11), or after a relapse (row 16). It should be noted that these information gaps surfaced only from the text in the related Reddit post. Further analysis of the comments associated with these posts can reveal additional information gaps and peer-suggested advice and self-treatment strategies to address these information gaps.

**Aspects of tapering buprenorphine products (TapeBup):** While analyzing posts that are only labeled with the theme *TapeBup*, we identified different aspects of tapering as well as different methods of tapering buprenorphine products (row 5). While analyzing posts that are labeled with other themes in addition to *TapeBup*, we identified several aspects of tapering related information needs. For instance, self-dosing and changing administration methods for tapering (row 11), seeking strategies to cope with psychophysical effects stemming from tapering, including using alternative treatments and controlled substances (rows 6, 10, 17, 18, 20).

**Co-occurring substance use while in recovery (CoSU):** Thematic analysis of posts labeled only with *CoSU* (row 4) reveals many affected individuals seek information about concurrent use of buprenorphine products and controlled substances (e.g., *Kratom, Shrooms, Alcohol*) and seek treatment options tailored to their substance dependence history (e.g., dependent on fentanyl patch vs. heroin). While analyzing posts labeled with other themes in addition to *CoSU*, we identified several information needs that can impact treatment induction and retention, e.g., how long to wait to start a buprenorphine product after substance use (row 8), use of controlled substances to cope with the psychophysical effects stemming from recovery treatment (row 9, 16, 19, 21, 24) or from tapering (rows 10, 17, 20, 22, 25).

### 3.3 Discovering Self-treatment Strategies

Our qualitative analysis also illuminates several self-treatment strategies for which individuals seek information from peers. These include asking questions about self-tapering buprenorphine products and self-dosing different medications to cope with the psychophysical effects of OUD treatment. Table 2 represents some example excerpts of seeking self-treatment strategies. The comments in these posts reveal peer-suggested self-treatment strategies for different themes.

*Table 2: Examples of self-treatment strategies*

| Self-treatment strategy | Examples |
|---|---|
| self tapering | *"Could someone suggest a tapering plan for quitting 4mg?..."*, *"..seeking advice on quitting subs using kratom…"* |
| self-dosing | *"... Is it okay to take 2mg of subs now?"*, *"... I only have a few 5mg hydrocodones. Can taking those help alleviate the current diarrhea and chills I'm experiencing?"* *"Clonadine can be helpful if you're getting withdrawal symptoms..."* |

### 3.4 Surfacing Rumors and Misperceptions

The individuals' inclination to seek advice from peers on various issues (row *'TapeBup'* and row *'CoSU-TakeBup'* in Table 4) inspired us to analyze some comments on the posts associated with these topics. Peers frequently offer advice based on their individual experiences, which can vary greatly from person to person, deviating from official guidelines. For example, even though it is advised not to take buprenorphine immediately after taking opioids (depending on the specific scenario, it is recommended to wait from 12-36 hours (ASAM, 2024)), a peer provided the following (paraphrased) response when asked how long to wait to take Subutex (a buprenorphine product) after using Oxycodone (an opioid). Thus, peers can inadvertently provide advice that is contrary to established clinical guidelines.

Peer suggestion: *You don't need to wait. I attempted the same approach a while back, and buprenorphine simply blocks the effects of other opioids.*

Our analysis also surfaced several rumors, i.e., suggestions that can not be clinically verified, e.g., tapering guidelines and regimen to quickly taper off buprenorphine products. Following is an example paraphrased post.

*I need help doing a rapid taper. 5 years ago, I was at 16mg and have tapered to 2mg. Within 5 days ago I am down to 1.5mg. How can I be done with it fast?*

Several comments in this post and similar posts contain peer suggestions on different aspects of tapering. Although these suggestions are often provided in good faith, they are challenging to verify clinically and are thus considered rumors. It should be noted that while not all rumors are harmful, some may contain potentially harmful information.

As another example, individuals might opt for alternative and unverified treatments, discontinue prescribed medications, or delay seeking professional help due to treatment misperceptions. Following is a paraphrased excerpt (Row 10 in Table 4) where an individual has decided to use kratom to taper down Suboxone by self-decision. However, Kratom is not clinically prescribed to be used during tapering buprenorphine products.

*Hello, seeking advice on quitting subs using kratom. Can I transition directly or should I taper off subs while starting kratom?*

In another post (Row 4 in Table 4), the individual intended to use 'shrooms (the slang/shorthand for Psilocybin Mushrooms, a controlled substance (DEA, 2024)) during the recovery as they found peers discussing the positive sides of this hallucinogen, although using such products is not recommended during MOUD treatment.

*Any insights on using 'shrooms while in recovery on Suboxone? It's often discussed as potentially beneficial for addicts.*

*Table 3: Frequency of the theme combinations. We used acronyms for each theme, such as AccBup: Accessing buprenorphine, CoSU: Co-occurring Substance Use, TakeBup: Taking buprenorphine, Psyphy: Experiencing Psychophysical Effects during Recovery, and TapeBup: Tapering buprenorphine. The ordering of themes inside the theme combinations is chronological and does not carry any positional significance.*

| Row No. | Theme(s) | Frequency | Percentage within the theme category | Percentage within the whole dataset |
|---|---|---|---|---|
| **Posts with a single theme (Total =2417, 40.3%)** | | | | |
| 1 | *Psyphy* | 700 | 29% | 11.7% |
| 2 | *AccBup* | 672 | 27.8% | 11.2% |
| 3 | *TakeBup* | 527 | 21.8% | 8.8% |
| 4 | *CoSU* | 281 | 11.6% | 4.7% |
| 5 | *TapeBup* | 237 | 9.8% | 4% |
| **Posts with two themes (Total = 2160, 36%)** | | | | |
| 6 | *Psyphy-TapeBup* | 738 | 34.2% | 12.3% |
| 7 | *TakeBup-Psyphy* | 391 | 18.1% | 6.5% |
| 8 | *CoSU-TakeBup* | 335 | 15.5% | 5.6% |
| 9 | *CoSU-Psyphy* | 274 | 12.7% | 4.6% |

| Row No. | Theme(s) | Frequency | Percentage within the theme category | Percentage within the whole dataset |
|---|---|---|---|---|
| 10 | *CoSU-TapeBup* | 105 | 4.9% | 1.8% |
| 11 | *TakeBup-TapeBup* | 105 | 4.9% | 1.8% |
| 12 | *AccBup-Psyphy* | 68 | 3.1% | 1.1% |
| 13 | *AccBup-CoSU* | 64 | 3% | 1.1% |
| 14 | *AccBup-TakeBup* | 57 | 2.6% | 1% |
| 15 | *AccBup-TapeBup* | 23 | 1.1% | 0.4% |
| **Posts with three themes (Total = 835, 13.9%)** | | | | |
| 16 | *CoSU-TakeBup-Psyphy* | 312 | 37.4% | 5.2% |
| 17 | *CoSU-Psyphy-TapeBup* | 233 | 27.9% | 3.9% |
| 18 | *TakeBup-Psyphy-TapeBup* | 142 | 17% | 2.4% |
| 19 | *AccBup-TakeBup-Psyphy* | 34 | 4.1% | 0.6% |
| 20 | *CoSU-TakeBup-TapeBup* | 33 | 4% | 0.6% |
| 21 | *AccBup-CoSU-Psyphy* | 31 | 3.7% | 0.5% |
| 22 | *AccBup-Psyphy-TapeBup* | 27 | 3.2% | 0.5% |
| 23 | *AccBup-TakeBup-TapeBup* | 8 | 1% | 0.1% |
| 24 | *AccBup-CoSU-Psyphy* | 7 | 1% | 0.1% |
| 25 | *AccBup-CoSU-TakeBup* | 7 | 0.8% | 0.1% |

## 4. Discussion

In this study, we applied our novel theme-driven framework to curate a large dataset coded with multiple themes and to perform a thorough qualitative analysis involving individual and co-occurring themes. Our analysis surfaces common information needs, knowledge gaps, and misperceptions that adversely impact initiation and adherence to OUD treatment using buprenorphine products. To the best of our knowledge, this is the first approach to characterize themes related to information seeking regarding buprenorphine products for OUD treatment from thousands of users. Our findings can complement traditional methods like patient surveys and interviews to understand critical gaps in OUD treatment delivery. Moreover, data-driven insights can help clinicians, clinical researchers, social workers, and policymakers achieve a better understanding of patients and improve patient education and communication.

**4.1 Inform Research on Patient Education and Patient-provider Communication**

Our results surface information gaps across different stages of treatment, such as treatment initiation, tapering, as well as different relevant events, e.g., co-occurring substance use, and the emergence of new treatment options (e.g., Brixadi). These findings can provide insights into effectively communicating with patients, providers, recovery coaches, and peer support providers about critical information gaps and how to address them. These findings can also inform approaches to proactively address concerns that might impact treatment induction, adherence, and retention, e.g., addressing concerns about tapering and long-term use of buprenorphine to avoid tapering and self-dosing, addressing information gaps about medication administration to reduce challenges associated with perceived psychophysical effects resulting from buprenorphine products. The findings can also inform when and why people deviate from clinical guidelines, thereby informing the design of tailored informational interventions to reduce the deviation.

**4.2 Designing Interventions to Address Treatment-related Misperceptions and Rumors**

As discussed in the results sections, there are several critical misconceptions and rumors related to treatment that can impact health safety and treatment outcomes, e.g., false risk perceptions about using different alternate treatments, medications, and controlled substances to cope with psychophysical effects, self-dosing buprenorphine products, concern about long term use of buprenorphine. Further research is needed to identify the prevalence of such misperceptions among different patient populations and geographic locations to design interventions to address these misperceptions and rumors. Another direction can be developing online informational interventions to mitigate the effect of potentially harmful information, as demonstrated in research on vaccine hesitancy and lack of adherence to clinical guidelines for infectious diseases.

**4.3 Generate Hypothesis for Clinical Research and Public Health Research**

Analyzing such a large-scale, patient-generated discourse can also enable clinical researchers to generate hypotheses, e.g., investigating the link between a prevalent self-reported perceived psychophysical effect and the corresponding buprenorphine product surfaced from the data, clinically supervised strategies to cope with withdrawal and other severe psychophysical effects, characterize the effectiveness of different tapering approaches based on patient's medical history (e.g., co-occurring mental health conditions) and substance use history, exploring the effectiveness of different treatment options based on patient's substance use history (e.g., effective treatment option for Xylazine vs. Fentanyl dependency). In addition, our data also reveal several cases to capture patients' lived experiences in a community-informed way to increase treatment adherence and retention, e.g., addressing concerns about coping with the psychophysical effects of treatment, long-term effects of treatment, and access barriers to treatment. This can inform the design of tailored patient-centric studies for public health and addiction researchers. Also, linguistic analysis of the peer supporter using the comments in the posts can generate insights to improve patient-provider communication and interaction with peer supporters and recovery coaches, as well as research on addressing stigma.

## 4.4 Complementing Existing Sampling-based Thematic Analyses

Our dataset and thematic analysis are valuable for analyzing different aspects of OUD treatment due to its large sample size. Although platforms like Reddit offer a valuable source of real-world spontaneous data for substance use disorder (SUD) research, the existing sampling-based thematic analyses cannot fully leverage the potential of the data. These analyses often employ a limited number of samples to identify prevalent topics, which often constrain the generalizability of their findings. The limited sample size used by these approaches also impedes the application of computational models to gather rich user-generated content from a broad demographic. In contrast, our extensive dataset, comprising 6,000 posts contributed by 3372 unique individuals, offers a robust resource for identifying frequently observed TINs related to OUD treatment using buprenorphine products. It provides a solid foundation for developing new natural language processing (NLP) models for OUD and SUD research in a community-informed way. Also, existing datasets are often not publicly accessible, limiting reproducibility. We will publicly release our dataset to promote future research following Reddit's data-sharing policies.

## 5. Conclusion

We focused on Reddit data, which is subject to gender and age biases - predominantly used by individuals between 18-49 (MarketingCharts, 2024a) and by male individuals (MarketingCharts, 2024b). Meanwhile, the list of themes we considered here is deemed important and significant by experts. However, it is not exhaustive, and there may be other noteworthy themes not included. An alternative conceptualization of the themes could also lead to different results. Again, a subreddit focusing on Methadone based OUD treatment might yield different themes. Moreover, our dataset cannot provide information on aspects that individuals do not self-disclose, such as using non-prescribed buprenorphine products for treatment. Additionally, we lack data on how many individuals are already in treatment or considering starting treatment unless they are self-disclosed.

Our primary purpose was to identify the treatment information needs (TINs) of individuals considering or undergoing OUD treatment using buprenorphine products on Reddit. The thematic analysis is a valuable resource for gaining deeper insights into increasing treatment induction, adherence, and retention while paying attention to patients' sense of autonomy and concerns about long-term treatment's safety, effectiveness, and accessibility. Overall, the curated dataset (i.e. original posts and comments sorted according to themes) can contribute to examining treatment safety, effectiveness, and accessibility for individuals with OUD. Moving forward, this work can be a basis for several potential future research studies. We can leverage the dataset and framework to gather insights on socio-cultural, behavioral, and health-related questions for minority health and health disparities following the NIMHD Research Framework (National Institute on Minority Health and Health Disparities, 2023). Although Reddit is an anonymous platform, such information is sometimes self-disclosed by individuals (Choudhury and De, 2014; Miller, 2020). While this paper mostly focuses on analyzing the posts, further analysis of the comments associated with these posts can reveal additional insights. Also, future research can use this framework in other subreddits and online platforms (e.g., YouTube, Facebook) to streamline the TIN identification process and potential interventions to address these TINs.

*Table 4: Observed frequently discussed topics for each theme. We used acronyms for each theme, such as AccBup=Accessing buprenorphine, CoSU=Co-occurring Substance Use, TakeBup= Taking buprenorphine, Psyphy= Experiencing Psychophysical Effects during Recovery, TapeBup=Tapering buprenorphine, AccBup-CoSU-TakeBup=Combination of Accessing buprenorphine - Co-occurring Substance Use - Taking buprenorphine. The ordering of themes inside the theme combinations is chronological and does not carry any positional significance.*

| Row No. | Theme | Commonly Discussed topics with the theme | Examples (paraphrased and redacted samples) |
|---|---|---|---|
| | **Posts with a single theme** | | |
| 1 | *Psyphy* | Seeking peer's experiences and advice on psychological or physical effects (e.g., tooth problem, shy bladder) experienced while undergoing treatment with buprenorphine products for OUD treatment. | *After 14 months on 6mg of subs, my teeth are constantly aching. Does anyone else experience this side effect?* |
| | | Seeking information on the reason (e.g., why the problem happens only in the morning, why the problem has arisen now albeit the same usage as before) of a particular side effect (e.g., morning sickness, headaches, agitated feeling) that emerged from the use of buprenorphine products. | *My significant other and I transitioned from fentanyl to Suboxone, and today marks our 30th day on it. We've noticed that while we feel fine during the day, both of us experience sneezing, belly cramps, and an unpleasant beginning-of-sickness feeling after a night's sleep. We haven't missed any doses. What could be causing this?* |
| 2 | *AccBup* | Seeking information on different aspects (e.g., how to enroll in the telemedicine service, how much time to wait to get the refill) of online medication providers (e.g., Quick.md, Bicycle Health, Bupe.me). | *I'm with Bicycle Health now, but I might get dismissed since I can't finish due to home tests. Can I still join Quick.md?* |
| | | Seeking information on Pharmacy refill issues (e.g., pharmacy discontinued the previous brand, delay in refill by the pharmacy). | *Could someone please inform me about big chain pharmacies that stock the Sandoz brand? The pharmacy I typically visit has recently switched to Alvogen.* |

| Row No. | Theme | Commonly Discussed topics with the theme | Examples (paraphrased and redacted samples) |
|---|---|---|---|
|  |  | Seeking information on insurance problems (e.g., sudden loss of insurance, whether a particular insurance covers a brand, pharmacy not accepting a particular insurance). | I'm wondering if anyone has information on whether Independent Health provides coverage for generic Suboxone brands? |
| 3 | TakeBup | Seeking information on the technique to administer (e.g., dissolving under the tongue, spitting, swallowing) the buprenorphine products. | Where exactly should I place the sub for the "gum and cheek" method? I'm tired of putting it under my tongue; it feels like half of it goes to waste. |
|  |  | Seeking information on the comparative discussion (e.g., which one absorbed better, which one is easy to dose) between two forms (e.g., films vs tablets), brands (e.g., name brand vs generic brand), or types (e.g., Suboxone vs Sublocade) of buprenorphine products. | Do the newer Suboxone pills work the same way as the strips or should they be taken like regular pills? I'm considering switching from strips to pills and wanted to clarify. |
| 4 | CoSU | Seeking information on the concurrent use of buprenorphine products and controlled substances (e.g., Kratom, Shrooms, Alcohol). | Any insights on using 'shrooms while in recovery on Suboxone? It's often discussed as potentially beneficial for addicts. |
|  |  | Seeking information on starting buprenorphine products to recover from current substance use (e.g., Fentanyl, Morphine). | I'm dealing with a significant Fentanyl dependency .... Has anyone here successfully used Suboxone to manage fentanyl addiction? If so, how did you go about it? |
| 5 | TapeBup | Seeking information on the appropriate technique (e.g., the most suitable lower dose to jump, the proper duration to stay on the lower dose before quitting, the best plan to taper given 50 tablets in hand) to taper buprenorphine products. | Could someone suggest a tapering plan for quitting 4mg? I've been on subs for 4-5 years. |
|  |  | Seeking peer's suggestion on a particular step or condition (e.g., the feasibility of jump offing at 0.4mg, tapering experience during pregnancy) | I've learned I'm pregnant and plan to taper my dose, but I'm worried about NAS or losing custody if I don't quit entirely. Has anyone given birth on Suboxone? What was your experience? |

| Row No. | Theme | Commonly Discussed topics with the theme | Examples (paraphrased and redacted samples) |
|---|---|---|---|
| | | of a specific tapering strategy (e.g., slow taper, fast taper). | |
| | **Posts with two themes** | | |
| 6 | *Psyphy-TapeBup* | Seeking information on physical or psychological effects (e.g., withdrawal, constipation, anxiety) while tapering or quitting buprenorphine products. | *I'm on day 9 of Suboxone withdrawal, feeling hazy. How much longer will this last? I quit at 4mg.* |
| 7 | *TakeBup-Psyphy* | Seeking information on changing the brand (e.g., name brand to generic brand) or type (e.g., Suboxone to Subutex) of buprenorphine products due to the side effects (e.g., feeling tired, affecting stomach) caused by the current brand or type. | *Subs helped me be present for my family, but they come with overwhelming depression, possibly due to naloxone. Does switching to Subutex or Zubsolv, known to be milder, improve this? Any personal experiences?* |
| 8 | *CoSU-TakeBup* | Seeking information on the proper time gap to switch from substance (e.g., Fentanyl, Heroin) use to buprenorphine products. | *I relapsed today. Can I take Subutex tomorrow, considering it hasn't fully left my system?* |
| 9 | *CoSU-Psyphy* | Seeking information on using substances (e.g., Alcohol, Oxycodone, Percocet) while on buprenorphine products, and the resultant side effects (e.g., feeling sick, feeling shit, withdrawal). | *Taking 2mg subs for 3 years, is it okay to have a glass of wine without getting sick?* |
| 10 | *CoSU-TapeBup* | Seeking information on using substances (Kratom, Imodium, etc.) during tapering a buprenorphine products. | *Hello, seeking advice on quitting subs using kratom. Can I transition directly or should I taper off subs while starting kratom?* |
| 11 | *TakeBup-TapeBup* | Seeking information on the technique to administer buprenorphine products (e.g., efficiency of diluting into the water, way to cut the strips perfectly) during tapering. | *A few mentioned volumetric dosing during sub-2mg taper. Can you explain this approach?* |
| 12 | *AccBup-Psyphy* | Seeking information on managing physical or psychological effects (e.g., withdrawal, feeling sick) due to running out early on buprenorphine | *A few days ago, I mentioned running out of my 6mg/day Subutex prescription. It's now around day 6, and I feel terrible. Can the ER assist? I* |

| Row No. | Theme | Commonly Discussed topics with the theme | Examples (paraphrased and redacted samples) |
|---|---|---|---|
| | | products for different reasons (e.g., taking extra doses, losing some tablets/pills). | ran out much earlier than expected, and I can't refill for a week. |
| 13 | *AccBup-CoSU* | Seeking information on getting a prescription/refill of buprenorphine products after relapse or recreational use of a specific substance (e.g., Kratom, Oxycodone, Benzos). | I'm concerned my doctor might discontinue my Subutex prescription if there are Ritalin and benzos in my system without his prescription. |
| | | Seeking induction strategies to start OUD treatment using buprenorphine products from specific substance dependency (e.g., Fentanyl, Oxycodone) | I must quit Kratom, I can't continue. Suboxone seems like the solution, and I need it urgently. Quick.md has positive reviews, but does anyone know if they prescribe Suboxone for Kratom addiction? |
| 14 | *AccBup-TakeBup* | Seeking information on the individuals' experience with the changed brand (e.g., name brand to generic brand) or changed medication (e.g., Suboxone to Zubsolv) due to the unavailability of the regularly used brand/medication for different reasons (e.g., pharmacy not storing the current brand, insurance does not cover the current brand/medication, the health provider changes the brand/medication) | Has anyone here experimented with or is presently using the Butrans patch? I'm contemplating using it since my insurance won't cover Belbuca. Does it adhere reliably for the full 7 days? How effective is it for pain relief? Thanks, everyone! |
| 15 | *AccBup-TapeBup* | Seeking information on a suitable tapering plan for a buprenorphine product use as they anticipate the next prescription refill for the buprenorphine product will be unavailable (e.g., due to the pharmacy discontinuing the medication brand, an individual does not have insurance/money to pay for the next refill). | Today, I collected my prescription which costs $283.96. I can't afford this amount. I think it's time to taper. I intend to reduce from 1.5 strips to 1 for 2 weeks, then to half, and so on. Is this tapering pace too rapid? |

| Row No. | Theme | Commonly Discussed topics with the theme | Examples (paraphrased and redacted samples) |
|---|---|---|---|
| | **Posts with three themes** | | |
| 16 | *CoSU-TakeBup-Psyphy* | Seeking information on a particular strategy (e.g., a particular time gap, a particular dosing schedule) to transition (for the first time or after a relapse) from substance use (e.g., Fentanyl, Heroin) to OUD treatment with buprenorphine products, and the corresponding side effects (e.g., withdrawal, sweating, stomach cramp). | *Took 2, 8mg strips daily for 3 days. Drank and used heroin last night. Wondering if I should wait 24+ hours to take Suboxone or if I can take it sooner to avoid withdrawal discomfort.* |
| 17 | *CoSU-Psyphy-TapeBup* | Seeking information on the use of substances (e.g., Kratom, Adderall, Clonazepam) to get rid of the side effects (e.g., withdrawal, body aches) that emerged while tapering or quitting buprenorphine products. | *Has anyone used kratom for Suboxone withdrawal? If yes, I'd appreciate hearing your experiences. I've been trying it during my first week off Suboxone, but I'm feeling uncertain about it.* |
| 18 | *TakeBup-Psyphy-TapeBup* | Seeking information on tapering or quitting one buprenorphine product due to its side effects (e.g., tiredness, feeling shitty) and switching to another buprenorphine product (e.g., Suboxone to Sublocade, Subutex to Suboxone). | *I've been taking 2-8mg of Suboxone since September. It's made me consistently tired and unmotivated. How have others successfully stopped using it? I've discussed sublocade with my doctor, and it seems like the best choice for me now.* |
| 19 | *AccBup-TakeBup-Psyphy* | Seeking information on the side effects (e.g., withdrawal, stomach pain, lethargy) caused by the change of buprenorphine brand (e.g., name brand to generic brand) or form (e.g., tablets to films), or medication type (e.g., Subutext to Suboxone) due to a logistic barrier (e.g., insurance issue, pharmacy supplying a different brand, intentional change by the health provider) | *I've been using Aquestive/Indivior 4mg twice daily, but the local pharmacy is shifting to Alvogen's generic strips. I've mostly found negative feedback about Alvogen. My primary worry is potential withdrawal symptoms. Can it cause precipitated withdrawal?* |
| 20 | *CoSU-TakeBu* | Seeking information on the proper time gap to start taking buprenorphine | *It's been around 8 hours since I stopped using kratom. Is it okay to take* |

| Row No. | Theme | Commonly Discussed topics with the theme | Examples (paraphrased and redacted samples) |
|---|---|---|---|
|  | p-Tape Bup | products for the first time after the use of substances (e.g., Kratom, Heroin), with an intent to use buprenorphine product for a short period of time and then taper it rapidly. | *2mg of subs now? I'm aiming for a two-week rapid taper to eventually stop everything.* |
| 21 | AccBup-CoSU-Psyphy | Seeking information on the use of substances (e.g., Oxycodone, Kratom, Hydrocodone) to reduce the side effects caused by the sudden stop of buprenorphine product use due to a logistic barrier (e.g., running out early, insurance problem). | *I've been on Suboxone for a very extended period, taking 24mg daily. Now, I'm completely out of it. I only have a few 5mg hydrocodones. Can taking those help alleviate the current diarrhea and chills I'm experiencing?* |
| 22 | AccBup-Psyphy-TapeBup | Seeking information on the physical or psychological side effects (e.g., withdrawal, anxiety, insomnia) while doing a taper due to a logistical barrier (e.g., running out early, insurance expiration, high medication cost, sudden unavailability of health provider). | *My boyfriend used to take 8mg of Suboxone daily for years. When he lost insurance a month ago, he had to quit. His last dose was around 6/20, and now it's 7/11, but he's still experiencing severe withdrawals - insomnia, skin discomfort, vomiting, and weakness.* |
| 23 | AccBup-TakeBup-TapeBup | Seeking information on the effective way (e.g., cutting medicine, volumetric dosing) to divide a high-strength medication (e.g., 8 mg strips) to a lower dose (e.g., 0.25 mg) while doing a taper due to a logistic barrier (e.g., insurance issue, run out early). | *I was given excessive 20mg daily dosage, abruptly stopped in the 5th month due to insurance issues. My plan was to stabilize and then gradually reduce by 0.25mg, with a week between reductions. I've heard that the buprenorphine distribution on the strips isn't uniform. Is volumetric dosing the better option?* |
| 24 | AccBup-CoSU-TapeBup | Seeking information on the use of substances (e.g., Kratom) while doing a taper due to a logistic barrier (e.g., insurance issue, run out early). | *My doctor stopped my auto-pay, leaving me with a high bill I can't afford right now. I'm on day 8 of quitting an 8mg three times a day medication cold turkey, and it's been incredibly challenging. I tried using kratom, but it hasn't provided any relief. Any suggestions?* |

| Row No. | Theme | Commonly Discussed topics with the theme | Examples (paraphrased and redacted samples) |
|---|---|---|---|
| 25 | *AccBup-CoSU-Take Bup* | Seeking information on the proper way (e.g., the proper time gap between the last substance use and the next buprenorphine dosing) to start the buprenorphine product after a forced relapse because of the temporary unavailability of the buprenorphine product due to a logistic barrier (e.g., running out early, insurance problem). | *After my final buprenorphine dose on Tuesday, I had to switch to Oxycodone due to the pharmacy's stock shortage. How much time must I wait before resuming the subs?* |

# 6. References

American Society of Addiction Medicine (ASAM). https://www.asam.org/docs/default-source/education-docs/unobserved-home-induction-patient-guide.pdf. Last accessed [July 22, 2024]

Bremer, W., Plaisance, K., Walker, D., Bonn, M., Love, J.S., Perrone, J., Sarker, A., 2023. Barriers to opioid use disorder treatment: A comparison of self-reported information from social media with barriers found in literature. Front. Public Health 11.

Centers for Disease Control and Prevention, 2023. CDC WONDER [online database]. Atlanta, GA: CDC. Available at: https://wonder.cdc.gov. [Last accessed: August 7, 2023]

Chen, J., Wang, Y., 2021. Social Media Use for Health Purposes: Systematic Review. J. Med. Internet Res. 23, e17917. https://doi.org/10.2196/17917

Choudhury, M.D., De, S., 2014. Mental Health Discourse on reddit: Self-Disclosure, Social Support, and Anonymity. Proc. Int. AAAI Conf. Web Soc. Media 8, 71–80. https://doi.org/10.1609/icwsm.v8i1.14526

Drug Enforcement Administration (DEA), 2024. Available at https://www.dea.gov/sites/default/files/2018-07/DIR-022-18.pdf. Last accessed [July 23, 2024]

Edo-Osagie, O., De La Iglesia, B., Lake, I., Edeghere, O., 2020. A scoping review of the use of Twitter for public health research. Comput. Biol. Med. 122, 103770. https://doi.org/10.1016/j.compbiomed.2020.103770

Jha, D., Singh, R., 2020. Analysis of associations between emotions and activities of drug users and their addiction recovery tendencies from social media posts using structural equation modeling. BMC Bioinformatics 21, 554. https://doi.org/10.1186/s12859-020-03893-9

Kanchan, S., Gaidhane, A., n.d. Social Media Role and Its Impact on Public Health: A Narrative Review. Cureus 15, e33737. https://doi.org/10.7759/cureus.33737

Larochelle, M.R., Bernson, D., Land, T., Stopka, T.J., Wang, N., Xuan, Z., Bagley, S.M., Liebschutz, J.M., Walley, A.Y., 2018. Medication for Opioid Use Disorder After Nonfatal Opioid Overdose and Association With Mortality: A Cohort Study. Ann. Intern. Med. 169, 137–145. https://doi.org/10.7326/M17-3107

Leech, A.A., McNeer, E., Stein, B.D., Richards, M.R., McElroy, T., Dupont, W.D., Patrick, S.W., 2023. County-level Factors and Treatment Access Among Insured Women With Opioid Use Disorder. Med. Care 61, 816–821. https://doi.org/10.1097/MLR.0000000000001867

Li, F., Jin, Y., Liu, W., Rawat, B.P.S., Cai, P., Yu, H., 2019. Fine-Tuning Bidirectional Encoder Representations From Transformers (BERT)–Based Models on Large-Scale Electronic Health Record Notes: An Empirical Study. JMIR Med. Inform. 7, e14830. https://doi.org/10.2196/14830

Livingston, N.A., Davenport, M., Head, M., Henke, R., LeBeau, L.S., Gibson, T.B., Banducci, A.N., Sarpong, A., Jayanthi, S., Roth, C., Camacho-Cook, J., Meng, F., Hyde, J., Mulvaney-Day, N., White, M., Chen, D.C., Stein, M.D., Weisberg, R., 2022. The impact of COVID-19 and rapid policy exemptions expanding on access to medication for opioid use disorder (MOUD): A nationwide Veterans Health Administration cohort study. Drug Alcohol Depend. 241, 109678. https://doi.org/10.1016/j.drugalcdep.2022.109678

Mackey, K., Veazie, S., Anderson, J., Bourne, D., Peterson, K., 2020. Barriers and Facilitators to the Use of Medications for Opioid Use Disorder: a Rapid Review. J. Gen. Intern. Med. 35, 954–963. https://doi.org/10.1007/s11606-020-06257-4

MarketingCharts, 2024a. (February 21, 2024). Percentage of U.S. adults who use Reddit as of September 2023, by age group [Graph]. In Statista. Retrieved July 22, 2024, from https://www.statista.com/statistics/261766/share-of-us-internet-users-who-use-reddit-by-age-group/

MarketingCharts., 2024b (February 21, 2024). Percentage of U.S. adults who use Reddit as of September 2023, by gender [Graph]. In Statista. Retrieved July 22, 2024, from https://www.statista.com/statistics/261765/share-of-us-internet-users-who-use-reddit-by-gender/

Marks, S. J., Pham, H., McCray, N., Palazzolo, J., Harrell, A., Lowe, J., Bachireddy, C., Guerra, L., Cunningham, P. J., & Barnes, A. J., 2023. Associations Between Patient-Reported Experiences with Opioid Use Disorder Treatment and Unmet Treatment Needs and Discontinuation Among Virginia Medicaid Members. Substance abuse, 44(3), 196–208. https://doi.org/10.1177/08897077231186218

Miller, B., 2020. Investigating Reddit Self-Disclosure and Confessions in Relation to Connectedness, Social Support, and Life Satisfaction. J. Soc. Media Soc. 9, 39–62.

National Institute on Drug Abuse. Available at https://nida.nih.gov/research-topics/commonly-used-drugs-charts. [Last accessed Oct 15, 2023]

National Institute on Minority Health and Health Disparities. Available at https://www.nimhd.nih.gov/about/overview/research-framework/nimhd-framework.html.

Last accessed [October 15, 2023]

Neely, S., Eldredge, C., Sanders, R., 2021. Health Information Seeking Behaviors on Social Media During the COVID-19 Pandemic Among American Social Networking Site Users: Survey Study. J. Med. Internet Res. 23, e29802. https://doi.org/10.2196/29802

Nobles, A.L., Johnson, D.C., Leas, E.C., Goodman-Meza, D., Zúñiga, M.L., Ziedonis, D., Strathdee, S.A., Ayers, J.W., 2021. Characterizing Self-Reports of Self-Identified Patient Experiences with Methadone Maintenance Treatment on an Online Community during COVID-19. Subst. Use Misuse 56, 2134–2140. https://doi.org/10.1080/10826084.2021.1972317

Pandrekar, S., Chen, X., Gopalkrishna, G., Srivastava, A., Saltz, M., Saltz, J., Wang, F., 2018. Social Media Based Analysis of Opioid Epidemic Using Reddit. AMIA. Annu. Symp. Proc. 2018, 867–876.

Paul, M.J., Sarker, A., Brownstein, J.S., Nikfarjam, A., Scotch, M., Smith, K.L., Gonzalez, G., 2016. SOCIAL MEDIA MINING FOR PUBLIC HEALTH MONITORING AND SURVEILLANCE, in: Biocomputing 2016. Presented at the Proceedings of the Pacific Symposium, WORLD SCIENTIFIC, Kohala Coast, Hawaii, USA, pp. 468–479. https://doi.org/10.1142/9789814749411_0043

Pizzicato, L.N., Hom, J.K., Sun, M., Johnson, C.C., Viner, K.M., 2020. Adherence to buprenorphine: An analysis of prescription drug monitoring program data. Drug Alcohol Depend. 216, 108317. https://doi.org/10.1016/j.drugalcdep.2020.108317

PRAW: The Python Reddit API Wrapper. Available at https://github.com/praw-dev/praw. [Last accessed October 15, 2023]

Pushshift Reddit API Documentation. Available at https://github.com/pushshift/api. [Last accessed October 15, 2023]

Ravndal, E., Amundsen, E.J., 2010. Mortality among drug users after discharge from inpatient treatment: An 8-year prospective study. Drug Alcohol Depend. 108, 65–69. https://doi.org/10.1016/j.drugalcdep.2009.11.008

Shulman, M., Wai, J. M., & Nunes, E. V. (2019). Buprenorphine Treatment for Opioid Use Disorder: An Overview. CNS drugs, 33(6), 567–580. https://doi.org/10.1007/s40263-019-00637-z

Silverstein, S.M., Daniulaityte, R., Miller, S.C., Martins, S.S., Carlson, R.G., 2020. On my own terms: Motivations for self-treating opioid-use disorder with non-prescribed buprenorphine. Drug Alcohol Depend. 210, 107958. https://doi.org/10.1016/j.drugalcdep.2020.107958

Skaik, R., Inkpen, D., 2021. Using Social Media for Mental Health Surveillance: A Review. ACM Comput. Surv. 53, 1–31. https://doi.org/10.1145/3422824

Sordo, L., Barrio, G., Bravo, M.J., Indave, B.I., Degenhardt, L., Wiessing, L., Ferri, M., Pastor-Barriuso, R., 2017. Mortality risk during and after opioid substitution treatment: systematic review and meta-analysis of cohort studies. BMJ j1550. https://doi.org/10.1136/bmj.j1550

Yarborough, B.J.H., Stumbo, S.P., McCarty, D., Mertens, J., Weisner, C., Green, C.A., 2016. Methadone, Buprenorphine and Preferences for Opioid Agonist Treatment: A Qualitative Analysis. Drug Alcohol Depend. 160, 112–118. https://doi.org/10.1016/j.drugalcdep.2015.12.031